\documentclass[aps,superscriptaddress,amsfonts,amssymb,amsmath,eqsecnum,twocolumn
]{revtex4}
\usepackage[dvips]{graphicx,color}
\usepackage{psfrag}
\usepackage[dvips,unicode, colorlinks, citecolor=blue, linkcolor=red, urlcolor=blue]{hyperref}

\usepackage{euler}

\allowdisplaybreaks

\begin{document}
\title{Accurate calculation of thermal noise in multilayer coating}
\author{Alexey Gurkovsky and Sergey Vyatchanin}
\affiliation{Faculty of Physics, Moscow State University, Moscow, Russia}
\date{\today
	}
\begin{abstract}
We derive accurate formulas for thermal fluctuations in multilayer interferometric coating taking into account light propagation inside the coating. In particular, we calculate the reflected wave phase as a function of small displacements of the boundaries between the layers using  transmission line model for interferometric coating and derive formula for spectral density of reflected phase in accordance with Fluctuation-Dissipation Theorem. We apply the developed approach for calculation of  the spectral density of coating Brownian noise.
\end{abstract}

\maketitle

%\section{Model of transmission line for multilayer coating}\label{DMTransLine}
\section{Introduction}

Thermal fluctuations in the mirror  are becoming significant noise sources in the second generation gravitational-wave antennae (Advanced LIGO, Advanced VIRGO, HF GEO, TAMA) \cite{aligo}. The pioneering articles on this issue dealt with Brownian fluctuations in {\em the body} of mirror \cite{95Raab, 98Levin, 98Bondu}. Later the importance of thermoelastic noise \cite{99PLAbgv} was realized especially for mirrors manufactured from sapphire. The physical reasons behind thermoelastic noise are provided by fundamental thermodynamic fluctuations of temperature manifesting through thermal expansion. The same reason produces the thermorefractive noise \cite{00PLAbgv}  through the mechanism of relation between the refractive index and temperature. These results were obtained for model of infinite test mass (the mirror was considered occupying semi-infinite elastic space). Later these results were generalized for the finite-size mirror \cite{98Bondu, 00Thorne}.
  
Soon the importance of thermal noise in the mirror {\em coating} was realized as the parameters of coating may differ considerably from the mirror parameters bulk. Thermoelastic noise in interferometric coating was calculated in \cite{03PLAbv,04Fejer}. Later, following direction of H.J.~Kimble \cite{08Kimble} the potential of partial compensation of thermoelastic and thermorefractive noise in coatings was explored \cite{08Evans, 08Gorm}. However, today the Brownian noise in interferometric coating produces the main contribution into the noise spectrum  \cite{02Harry,06Harry,09Somiya} of gravitational wave antennae because the loss angles of substances used in coating (as usual, pair $Ta_2O_5,\ Si\,O_2$) are much greater than the loss angles in the same bulk materials. %The calculations of thermal noise in a finite-size mirror (both for coating and substrate) were recently accomplished \cite{}.

As a rule the calculation of Brownian and thermoelastic noise in coating is reduced to calculation of fluctuations of the total coating thickness. Strictly speaking, the light partially travels inside the coating, however, its power exponentially decreases with the depth when the light travels inside the coating. It allows considering the light to be perfectly reflected from the outer surface of the coating. Usually, the calculation of thermal noise in coating is provided under assumption that light is reflected from the front surface of the coating {\em without} detailed analysis of the light propagating inside the coating \cite{03PLAbv, 04Fejer, 02Harry, 06Harry, 08Evans, 08Gorm,09Somiya}.

 \begin{figure}[ht]
 \psfrag{rho}{$\rho$}
 \psfrag{rho1}{$\rho_1$}
 \psfrag{rho2}{$\rho_2$}
 \psfrag{zc}{$z_c$}
 \psfrag{z}{$z$} \psfrag{z0}{$z_0$}
 \psfrag{z1}{$z_1$} \psfrag{z2}{$z_2$} 
 \psfrag{zM}{$z_M$} \psfrag{zM+1}{$z_{M+1}$}
 \psfrag{zN}{$z_N$} \psfrag{zN-2}{$z_{N-2}$}
 \psfrag{zN+1}{$z_{N+1}$}
 \psfrag{r}{$\rho_o$}
 \psfrag{rn}{$\rho_o=\frac {1}{n_o}$}
 \psfrag{cap}{cap}
 \includegraphics[width=0.45\textwidth]{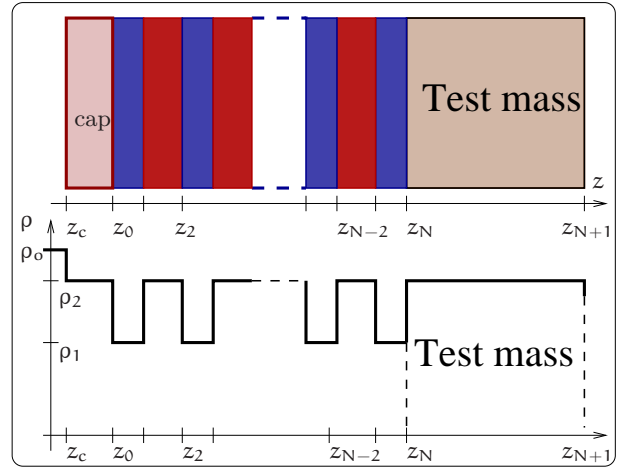}
 \caption{Top: Conventional mirror with interferometric coating deposited on the face side. The coating consists of alternating quarter wavelength layers plus the outer half wavelength cap layer.
 Bottom: transmission line model with alternating quarter wavelength parts, 
 $\rho_1=\frac{1}{n_1}\, \ \rho_2=\frac{1}{n_2},\ \rho_o=1$.}\label{CM}
\end{figure}

The purpose of this article is to make accurate calculations of reflected wave phase taking into account propagation of light inside the coating. In particular, we calculate the reflected wave phase as a function of small displacements of the boundaries between the layers. In our analysis we use  transmission line model for interferometric coating \cite{86Solimeno}. 
In Sec.~\ref{Calc} we calculate the dependence of the phase of the wave reflected from the mirror as function of small displacements of boundaries between interferometric layers of coating. In Sec.~\ref{Brownian} we apply the formula obtained to calculate the spectral density of Brownian noise inside the coating to obtain numerical estimates. In Sec.~\ref{Diss} we discuss obtained results.

\section{Calculation of reflected wave phase} \label{Calc}

It is known that the interferometric coating consists of alternating quarter wave length layers with different refraction indices $n_1, \ n_2$ ($n_1 >n_2$) as shown on Fig.~\ref{CM}. Specifically, below we assume that coating consists of $Ta_2O_5$ ($n_1$) and fused silica ($n_2$) layers, the outer layer (with positions $z_0,\ z_1$) is made from $Ta_2O_5$ and it is covered by additional half wave length layer (cap) which outer position is $z_c$. Such a coating is similar to transmission line consisting of alternating quarter wave length pieces with characteristic (wave) impedances $\rho_1=1/n_1$ and $\rho_2=1/n_2$ \cite{86Solimeno}.

We assume that the test mass (the body of the mirror) is manufactured from fused silica, hence, the wave impedance between $z_N$ and $z_{N+1}$ is equal to   $\rho_2=1/n_2$. 
The mean position $z_i$ of the surface of each layer (see Fig.~\ref{CM}) corresponds to the quarter wave length thickness of each layer, however, small position fluctuations $\zeta_i$ (caused by thermal fluctuations) produce fluctuations in reflected wave phase. In this section we calculate the reflected wave phase as a function of small displacements $\zeta_i$ using the successive  approximation technique.

\subsection{Main formulas and the zeroth approximation}

 In our consideration we ignore the optical losses in materials of coating.
We start with considering the last layer assuming that all fluctuation displacements $\zeta_i$ are equal to zero.  We assume that the wave propagating in the positive direction of axis $z$ is described as  $\sim e^{-i\omega(t-z/c)}$. The equivalent impedance at position $z=z_{N+1}$ is equal to $Z_{N+1}=\rho_o= 1$ (vacuum). It is convenient to define the amplitude reflectivy $R$ as
\begin{align*}
U_\text{refl}=R\,U_\text{incident} 
\end{align*}
Obviously, the amplitude reflectivity $R_{N+1}$ at $z=z_{N+1}$ is equal to:
\begin{align}
 Z_{N+1}&=\rho_o,\quad R_{N+1} = \frac{\rho_o-\rho_2}{\rho_2+\rho_o}
\end{align}
Now we can calculate the effective impedance $Z_{N}$ at position $z_{N}$ --- it allows considering the piece of transmission line between $z_N$ and $z_{N+1}$ as a single impedance $Z_{N}$
\begin{align}
 Z_N &= \rho_2\, \frac{1+R_{N+1}\theta_{N+1}^2}{1-R_{N+1}\theta_{N+1}^2},\quad 
 	Z_N^{(0)}=\frac{\rho_2^2}{\rho_o},\\
 \theta_{N+1}^2&\equiv \exp\big[2ik_{2}(z_{N+1}-z_{N})\big]=
 	-\exp\big[2ik_{2}(\zeta_{N+1}-\zeta_{N})\big],\nonumber\\ 
 	k_{1,2}&=k\, \frac{\rho_o}{\rho_{1,2}},\quad k=\frac{2\pi}{\lambda}\;.
\end{align}
Here $\theta_{N+1}$ is the exponent describing phase advance of wave traveling between positions $z_N$ and $z_{N+1}$, $k_{1,2}$ are the wave vectors, $\lambda$ is the optical wavelength in vacuum. Here and below superscript $^{(0)}$ refers to the particular case of the zeroth approximation when fluctuations $\zeta_N,\ \zeta_{N+1}$ are absent.   We  consider the test mass as  one fused silica layer (``slightly'' thicker). For convenience we assume its thickness being a fold to the quarter wave length (to meet the condition of anti resonance) --- it means $\theta_{N+1}^2=-1$. We make this assumption to make final formula more compact, without it the final result does not change practically but  formulas looks more akward.

In the similar manner we successively calculate reflectivity $R_N$, impedance $Z_{N-1}$ and so on:
\begin{subequations}
\begin{align}
 R_N &= \frac{Z_{N}-\rho_1}{Z_{N}+\rho_1},\quad R_N^{(0)}=
 	\frac{\frac{\rho_2^2}{\rho_1\rho_o}-1}{1+\frac{\rho_2^2}{\rho_1\rho_o}}\\
 Z_{N-1} & =\rho_1\, \frac{1+R_N\theta_N^2}{1-R_N\theta_N^2},\quad
 	Z_{N-1}^{(0)}=\frac{\rho_2^2}{\rho_o}=
 	\rho_o\,\frac{\rho_1^2}{\rho_2^2},\\
  \theta_{N}^2&\equiv  	-\exp\big[2ik_{1}(\zeta_{N}-\zeta_{N-1})\big],\\	
 R_{N-1} & =\frac{Z_{N-1}-\rho_2}{Z_{N-1}+\rho_2},\quad R_{N-1}^{(0)}=
        \frac{\rho_1^2\rho_o/\rho_2^3-1}{1+\rho_1^2\rho_o/\rho_2^3},\\
 Z_{N-2} & = \rho_2\, \frac{1+ R_{N-1}\theta_{N-1}^2}{1- R_{N-1}\theta_{N-1}^2},\quad
 	Z_{N-2}^{(0)}=   \frac{\rho_2^2}{\rho_o}\left(\frac{\rho_2^2}{\rho_1^2}\right),\\
 R_{N-2} &= \frac{Z_{N-2}-\rho_1}{Z_{N-2}+\rho_1},\quad
 	R_{N-2}^{(0)}=\frac{\frac{\rho_2^4}{\rho_1^3\rho_o}-1}{1+\frac{\rho_2^4}{\rho_1^3\rho_o}}\, ,
 	\quad \dots
\end{align}
\end{subequations}
%The last formula means that we can put single impedance $Z_ {N-2}$ at position $z_{N-2}$ instead of three last segments of transmission line (two layers and test mass). 
The total number $N$ of layers is assumed to be an odd number. Hence, we can calculate the impedance at point $z=z_0$ in the zero approximation:
\begin{align}
\label{Z0}
 Z_0^{(0)} & = \rho_o\left(\frac{\rho_1^2}{\rho_2^2}\right)^{(N+1)/2}=
 	\rho_o  \left(\frac{\rho_1}{\rho_2}\right)^{N+1}\ll \rho_o \, ,\\
 Z_{2m}^{(0)} &= \rho_o\left(\frac{\rho_1}{\rho_2}\right)^{N+1-2m},\quad
 Z_{2m-1}^{(0)} =\frac{\rho_2^2}{Z_{2m}^{(0)}}
\end{align}
Accounting for the half wave length cap layer gives that the effective impedance $Z_c^{(0)}$ at position $z_c$ is equal to $Z_c^{(0)}=Z_0^{(0)}$.

\paragraph*{Reflectivity and transparency.}
In case of small impedance $Z_0 \ll \rho_o$ one may easily calculate the effective amplitude reflectivity $R_0$ and transparency $T_0=\sqrt{1-R_0^2}$
\begin{align}
 R_0& =\frac{Z_0-\rho_o}{\rho_o+Z_0}\simeq -1 +\frac{2Z_0}{\rho_o},\quad
 	T_0 \simeq \frac{4Z_0}{\rho_o}
\end{align}
In the zeroth approximation we have:
\begin{align}
 R_0^{(0)} &= \frac{Z_0^{(0)}-\rho_o}{\rho_o+Z_0^{(0)}} = 
 -\,\frac{1-\left(\frac{\rho_1}{\rho_2}\right)^{N+1}}{1+\left(\frac{\rho_1}{\rho_2}\right)^{N+1}}
 	\,,\\
  \big(T_0^{(0)}\big)^2 &=
 \frac{4\left(\frac{n_2}{n_1}\right)^{N+1}}{
 \left(1+\left(\frac{n_2}{n_1}\right)^{N+1}\right)^2}\simeq 4\left(\frac{n_2}{n_1}\right)^{N+1}
 \end{align}

\subsection{Fluctuations of layer positions in coating}\label{Fluct}

Now we  take into account displacements $\zeta_j$ in position of each layer.  
We are interested in only linear terms of expansion $Z_0$ over $\zeta_j$. Hence, we can calculate each term separately, i.e. we can calculate a term, for example, proportional to a certain fluctuation $\zeta_m$ putting all other positions to be equal to zero: $\zeta_{i\ne m}=0$.  

\begin{figure}[ht]
 \psfrag{rho1}{$\rho_1$}
 \psfrag{rho2}{$\rho_2$}
 \psfrag{z}{$z$} \psfrag{z0}{$z_0$}
 \psfrag{z0+}{$z_0+\zeta_0$} 
 \psfrag{Z}{$Z_1^{(0)}$}
 \psfrag{rho0}{$\rho_o$}
 \psfrag{rn}{$\rho_o=\frac {1}{n_o}$}
 \includegraphics[width=0.25\textwidth]{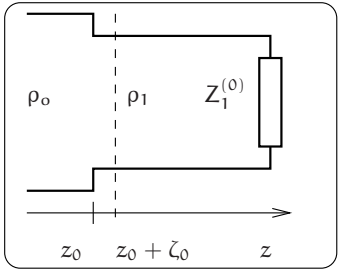}
 \caption{For calculation of the dependence on displacement $\zeta_0$.}\label{zeta0fig}
\end{figure} 
 
First we assume that the cap layer is absent (later we will include it into considerations). We start by taking into account a small displacement $\zeta_0$ assuming that all other displacements are equal to zero: $\zeta_{j\ne 0}=0$. Hence, we can account for the unperturbed piece of transmission line between $z_1\dots z_{N+1}$ as a single impedance $Z_{1}^{(0)}$ --- see Fig.~\ref{zeta0fig}. 

The displacement $\zeta_0$ influences the phase of reflected wave in two ways: a) through variation of effective impedance which, in turn, changes the reflectivity and b) through  direct displacement of front surface which reflects the wave (see formula (\ref{refl}) below). 
 
{\em a).} We calculate the perturbed impedance $Z_0$:
\begin{align}
 Z_1^{(0)} &= \frac{\rho_2^2}{Z_2^{(0)}}= \frac{\rho_2^2}{\rho_o} 
 	\left(\frac{\rho_2}{\rho_1}\right)^{N-1}\gg \rho_o,\\
 R_1^{(0)} &= \frac{Z_1^{(0)}-\rho_1}{Z_1^{(0)}+\rho_1}=\frac{1-\alpha_{N}}{1+\alpha_{N}},\quad
 	\alpha_{N}\equiv \frac{\rho_o}{\rho_2}
 	\left(\frac{\rho_1}{\rho_2}\right)^{N}\ll 1,\nonumber\\
 \theta_1^2 &\simeq -\big[1 -2ik_1\zeta_0\big],\\
 Z_0 &= \rho_1\frac{1+R_1^{(0)} \theta_1^2}{1-R_1^{(0)} \theta_1^2} =
 	\rho_1\frac{\alpha_{N} -(1-\alpha_{N})2ik_1\zeta_0 }{1+(1-\alpha_N)2ik_1\zeta_0 }\simeq\nonumber\\
 &= Z_0^{(0)}+
 	ik_1\zeta_0 \rho_1(1-\alpha_{N}^2)=\nonumber\\
 \label{zeta0}
 &=	Z_0^{(0)}+ ik\zeta_0 \rho_0\left(1-\left[
 		\frac{\rho_o}{\rho_2}\left(\frac{\rho_1}{\rho_2}\right)^{N}\right]^2\right).
\end{align}
Now we can estimate the perturbed reflectivity $R_0$:
\begin{align}
\label{R0}
 R_0 &=\frac{Z_0-\rho_o}{Z_0+\rho_o}\simeq R_0^{(0)}\left(1-
 	\frac{2ik\zeta_0(1-\alpha_{N}^2)}{1-\alpha_{N+1}^2}\right)\\
 	&\alpha_{N+1}\equiv \left(\frac{\rho_1}{\rho_2}\right)^{N+1}\ll 1  
\end{align}

{\em b).} Let the amplitude of the incident wave to be $A$ then the complex amplitude $B$ of the reflected wave is provided as 
\begin{align}
 \label{refl}
 B & = R_0 A\, e^{2ik\zeta_0}
 \end{align}
 Now expanding exponent in (\ref{refl}) in series over $\zeta_0$ and  substituting (\ref{R0}) we obtain
 \begin{align}
 B&\simeq
 	 R_0^{(0)}A\left(1-	\frac{2ik\xi_0(1-\alpha_{N}^2)}{1-\alpha_{N+1}^2}+2ik\zeta_0\right)\simeq
 	\\
 &\simeq R_0^{(0)}A\left(1-2ik\zeta_0 \left(\frac{\rho_1}{\rho_2}\right)^{2N}\frac{\rho_2^2-\rho_1^2}{\rho_2^2}\right).\nonumber
\end{align}
We see that contribution of displacement $\zeta_0$ into the phase of reflected wave is depressed by a small factor $\sim (\rho_1/\rho_2)^N$.

In the same manner we can calculate contribution of fluctuation displacement in each layer $\zeta_i$ into the phase of reflected wave. % including the displacement $\zeta_c$ of the front cap surface.
In addition we can take into account fluctuations of position $\zeta_c$ of additional half wavelength layer (cap) made from fused silica. Then we can calculate the phase of the reflected wave expressing it as a function of small displacements $\zeta_i$ with account of the cap  --- see detailed calculations in Appendix~\ref{AppA} 

\begin{figure}[ht]
 \psfrag{e}{$\epsilon_j$}
 \psfrag{rho2}{$\rho_2$}
 \psfrag{z}{$z$} \psfrag{z4}{$z_4$}
 \psfrag{z4+}{$z_4+\zeta_2$} 
 \psfrag{Z}{$Z_5^{(0)}$}
 \psfrag{rho0}{$\rho_o$}
 %\psfrag{rn}{$\rho_o=\frac {1}{n_o}$}
 \includegraphics[height=0.3\textwidth,width=0.4\textwidth]{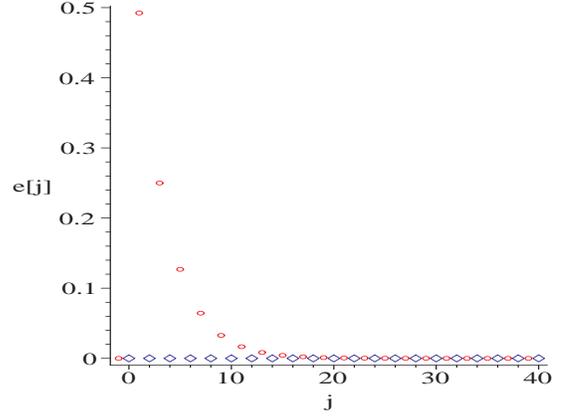}
 \caption{The coefficients $\epsilon_j$ from formula (\ref{epsiloncap}) for a coating with a cap. Red circles correspond to odd subscript and blue diamonds -- to even ones. $n_1=2.035\ (Ta_2O_5), n_2=1.45 \ (SiO_2)$. The number of $Ta_2O_5$ layers  is assumed to be $N_1=20$, and the number of $SiO_2$ layers --- $N_2=19$ plus cap.  Subscript $_{-1}$ on the plot corresponds to $\epsilon_c$. }\label{EpsilonCap}
\end{figure} 

The formula for phase $\phi$ of reflected wave  is convenient to write through refractive indices ($\rho_1=1/n_1,\ \rho_2=1/n_2$, $n_1>n_2$):
\begin{subequations}
\label{epsiloncap}
 \begin{align}
 i\phi&= 2ik\left(\epsilon_c\zeta_c+\sum_{j=0}^{N+1}\epsilon_j\, \zeta_j\right),\\
 \epsilon_c &= n_2\left(\frac{n_2}{n_1}\right)^{2N+2},\\
 \epsilon_0& = \left(\frac{n_1^2-n_2^2}{n_1^2} \right)n_2^2
   \left(\frac{n_2}{n_1}\right)^{2N}, \\ 
 \epsilon_1& = 
   \left(\frac{n_1^2-n_2^2}{n_1^2}\right),\\
 \epsilon_2& = \left(\frac{n_1^2-n_2^2}{n_1^2} \right)n_2^2
   \left(\frac{n_2}{n_1}\right)^{2N-2},\\
 \epsilon_3& = \left(\frac{n_1^2-n_2^2}{n_1^2}\right)
   \left(\frac{n_2}{n_1}\right)^2 ,\\
 \epsilon_4& =  \left(\frac{n_1^2-n_2^2}{n_1^2}\right) n_2^2 
   \left(\frac{n_2}{n_1}\right)^{2N-4},\dots \\
 \epsilon_{2m-1}& = \left(\frac{n_1^2-n_2^2}{n_1^2}\right)
   \left(\frac{n_2}{n_1}\right)^{2 m-2},\\
 \epsilon_{2m}& =  \left(\frac{n_1^2-n_2^2}{n_1^2}\right) n_2^{2} 
   \left(\frac{n_2}{n_1}\right)^{2N-2m},\dots\\
 \epsilon_{N}& = \left(\frac{n_1^2-n_2^2}{n_1^2}\right)
   \left(\frac{n_2}{n_1}\right)^{N-1},\\
 \epsilon_{N+1}& =  \left(\frac{n_1^2-n_2^2}{n_1^2}\right) n_2^{2} 
   \left(\frac{n_2}{n_1}\right)^{N-1}.
\end{align}
\end{subequations}
We see that  fluctuation displacements $\zeta_1,\ \zeta_3,\dots$ (odd numbers) provide the main contribution into sum (\ref{epsiloncap}) whereas the input of displacements $\zeta_c,\ \zeta_0,\ \zeta_2,\ \zeta_4\dots$ (even numbers) is negligible. The plot of coefficients $\epsilon_i$ is presented in Fig.\ref{EpsilonCap}. 

Note that formulas for coefficients $\epsilon_i$ obtained in the cases with and without half wavelength cap are practically the same  --- compare formulas (\ref{phi}) and plots in Fig.~\ref{Epsilon} in Appendix \ref{AppA} with formulas (\ref{epsiloncap}) and plot  in Fig.~\ref{EpsilonCap} correspondingly.

\section{Calculation of coating Brownian noise}\label{Brownian}

We can apply the exact formulas (\ref{epsiloncap}) to calculate Brownian noise to coating. In accordance with Fluctuation Dissipation Theorem (FDT) \cite{LL5, 51Callen, 98Levin}, in order to calculate the spectral density $S_X(\omega)$ of fluctuations of variable $X=\sum\epsilon_i\zeta_i$ at the frequency $\omega$,  we have to apply force $\sum_i\epsilon_i F_0e^{i\omega t}$ acting at frequency $\omega$ so that the force $\epsilon_i F_0$ is to be applied to position $z_i$. Then we have to calculate the total dissipated power $W$ and to find the spectral density using formula
\begin{align}
\label{FDT}
 S_X(\omega) = \frac{8k_BT\, W}{F_0^2\omega^2}
\end{align}
In so doing we just calculate the spectral density of fluctuating variable $X$ as there is no need to calculate additional correlations between thickness fluctuations in different layers. 

In contrast to previous approximate approaches  \cite{03PLAbv,04Fejer, 02Harry, 06Harry, 08Evans, 08Gorm} the formulas (\ref{epsiloncap}) provide the option of direct explicit calculation of the thermal noise in coating. Recall that in above articles the following approximate formula was used for reflected wave phase 
\begin{equation}
\label{phiApp}
 \phi_\text{approx}= 2ik\zeta_c
\end{equation}
In this case we have to apply force $F_0e^{i\omega t}$ to position $z_c$ in order to calculate spectral density of variable $\zeta_c$.

Obviously, the use of explicit formula (\ref{epsiloncap}) will give {\em smaller} value of dissipated power relative to result of approximate approach ({\ref{phiApp}). Hence, the spectral density should be also smaller.

\subsection{Mirror as an infinite half space}
We compare the explicit and approximate calculations of Brownian noise of coating using the model of the mirror as a semi-infinite half space using approach developed for structural losses in \cite{02Harry,06Harry}. In accordance with FDT \cite{51Callen,LL5, 98Levin} we have to apply pressure $\epsilon_i p_0$ to the $i$-th boundary between layers:
\begin{align}
 \epsilon_c p_0 &\quad \text{-- to outer surface of cap},\quad 
  p_0=\frac{2F}{\pi w^2}\, e^{-2r^2/w^2},\nonumber\\
  \epsilon_0 p_0 &\quad \text{-- to position }z_0,\\
  \epsilon_1 p_0 &\quad \text{-- to position }z_1,\ \dots
\end{align}
and so on. We are interested in elastic energy $U_i$ stored in each layer. 

First we write down formulas for strains $u_{ij}$ and stress $\sigma_{zz}$ in outer layer of the mirror body (substrate) \cite{98Bondu, 02Harry,LL}
\begin{subequations}
\begin{align}
 u_{rr} &= \frac{F\Sigma}{4\pi(\lambda+\mu)}\left(
  \frac{1}{r^2}\left[1-e^{-2r^2/w^2}\right]- 
    \frac{4}{w^2}e^{-2r^2/w^2}\right),\\
 u_{\phi\phi} &= \frac{-F\Sigma}{4\pi(\lambda+\mu)}
  \frac{1}{r^2}\left[1-e^{-2r^2/w^2}\right],\\
 u_{zz} &= \frac{-F\Sigma}{4\pi(\lambda+\mu)}
    \frac{4}{w^2}e^{-2r^2/w^2},\quad \\
  & \Rightarrow\ u_{zz}= u_{rr}+u_{\phi\phi},\nonumber\\
 u_{rz} &=0,\quad \Sigma \equiv \epsilon_c +\sum_{i=0}^N\epsilon_i,\\
 \sigma_{zz} &=\frac{{2}F\Sigma}{\pi w^2}\, e^{-2r^2/w^2}.  
\end{align}
\end{subequations}
Here $\lambda,\ \mu$ are the Lam\'e coefficients of substrate, which are known to be expressed through the Young modulus $Y$ and the Poisson ratio $\nu$ as following
\begin{align}
 \lambda &\equiv \frac{\nu Y}{(1+\nu)(1-2\nu)},\quad \mu\equiv \frac{Y}{2(1+\nu)},
\end{align}
We assume that the number $i=N$ corresponds to the surface of substrate (as  shown in Fig.~\ref{CM}).. 

Now we calculate strains and stresses in coating layers. As usual we assume that the tangent strains in the $i$-th layer of coating (the positions of its boundaries are $z_{i-1},\ z_i$) are equal to corresponding strains in substrate:
\begin{align}
 u_{rr}^{i} &=u_{rr},\quad u_{\phi\phi}^{i}=u_{\phi\phi} ,\quad
  u_{rz}^{i} =u_{rz}=0,
\end{align}
The last components $u_{zz}^{i}$ of the normal strain in the $i$-th layer can be found from the known formula binding stress and strain tensors \cite{LL}:
\begin{align}
\label{sigmazzi}
 \sigma_{zz}^{i} = (\lambda_{i} +2\mu_{i})u_{zz}^{i} +
  \lambda_{i}\big(u_{rr}^{i}+u_{\phi\phi}^{i}\big)
\end{align}
The normal stress $\sigma_{zz,\ i}^{i}$ can be easily calculated as following
\begin{align}
\label{sigmazz}
 \sigma_{zz}^{i} &= \frac{F\Sigma_i}{\pi w^2}\, e^{-2r^2/w^2},\quad
  \Sigma_i \equiv\epsilon_c +\sum_{j=0}^{i-1}\epsilon_j.
\end{align}
% One can write (\ref{sigmazzi}) for substrate
% \begin{align}
%  \sigma_{zz} & = (\lambda +2\mu) u_{zz} +\lambda\big(u_{rr}+u_{\phi\phi}\big)=
%   2(\lambda +\mu) u_{zz},\\
%   \sigma_{zz} &\equiv \sigma_{zz}^{N+1} ,\quad \Sigma\equiv \Sigma_{N+1}.
% \end{align}

\begin{figure}[ht]
 \psfrag{V[j]}{$V_i$}
 \psfrag{j}{$i$}
 \psfrag{z}{$z$} \psfrag{z4}{$z_4$}
 \psfrag{z4+}{$z_4+\zeta_2$} 
 \psfrag{Z}{$Z_5^{(0)}$}
 \psfrag{rho0}{$\rho_o$}
 %\psfrag{rn}{$\rho_o=\frac {1}{n_o}$}
 \includegraphics[height=0.3\textwidth,width=0.4\textwidth]{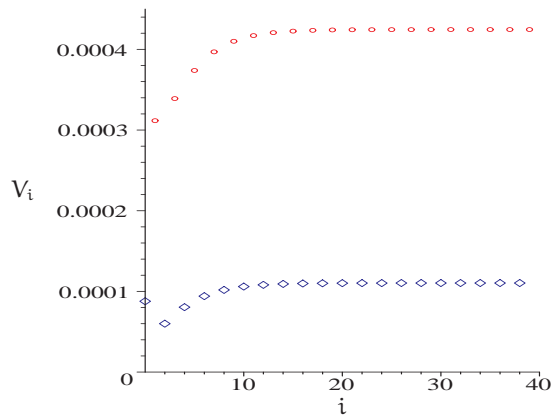}
 \caption{The coefficient $V_i$ proportional to energy stored in the $i$-th layer from formula (\ref{epsiloncap}) for coating with a cap. Red circles correspond to odd subscript ($n_\text{odd}=2.035\ (Ta_2O_5)$) and blue diamonds ($n_\text{even}=1.45 \ (SiO_2)$) -- to even ones. Number of $Ta_2O_5$ layers  is assumed to be $N_1=20$, and the number of $SiO_2$ layers --- $N_2=19$. Coefficent $V_{0}$ on the plot corresponds to the cap layer.}\label{EpsilonCapU}
\end{figure} 

 Manipulating with (\ref{sigmazzi}, \ref{sigmazz}) we  write down a useful formula:
\begin{align}
 u_{zz}^i &= A_i u_{zz}, \quad 
 A_i= \frac{2(\lambda +\mu)\Sigma_i-\lambda_i\Sigma}{(\lambda_i+2\mu_i)\Sigma}
\end{align}
and to calculate the elastic energy $U_i$ in the  $i$-th layer with thickness $d_i$ through its volume density $w_i$ using the known formulas:
\begin{align}
 w_i &= \mu_i\big(u_{rr}^2+u_{\phi\phi}^2+  (u_{zz}^i)^2\big)+
  \frac{\lambda_i}{2}\big(u_{zz}^i+u_{rr}+u_{\phi\phi}\big)^2, \nonumber\\
%  & = \mu_i\big(u_{rr}^2+u_{\phi\phi}^2 +  (u_{zz}^i)^2\big)+
%  \frac{\lambda_i}{2}\big(u_{zz}^i+u_{zz}\big)^2,\\
  U_i &= d_i\, 2\pi \int_0^\infty w_i\, r\, dr=\nonumber\\
  \label{Ui}
   = & \frac{d_i}{8\pi} \left(\frac{F\Sigma_N}{w(\lambda+\mu)}\right)^2
  \left[2\mu_i \big(1 +A_i^2\big)+
  \lambda_i \big(1+A_i\big)^2\right].
\end{align}

Now we calculate the spectral density of variable $X$ (\ref{FDT}) using the fact that for the structural losses the power dissipated in the $i$-th layer can be expressed as  $W_i=U_i\phi_i\omega$:
\begin{align}
  S_X(\omega) &=	\frac{k_B T}{\omega}\frac{\Lambda}{\pi(\mu+\lambda)}
   \left(\frac{F\Sigma_N}{2w}\right)^2\sum_{i=0}^N V_i,\\
  \label{Vi}
  V_i&=\left[1+\delta_{0i}\right]\frac{2\mu_i \big(1 +A_i^2\big)+  \lambda_i \big(1+A_i\big)^2}{n_i(\mu+\lambda)}\, \phi_i
\end{align}
Here  $\phi_i$ is the loss angle of structural losses in the  $i$-th layer, $\Lambda$ denotes the optical wavelength in vacuum, thickness $d_i$ of $i$-th layer is expressed through its refractive index $n_i$ as $d_i=\Lambda/4n_i$ for each layer except cap, the multiplier $\left[1+\delta_{0i}\right]$ is introduced to account for the fact that the cap thickness is two times larger (as its width is equal to {\em half} wavelength). The plot of dimensionless coefficients  $V_i$ is presented on Fig.~\ref{EpsilonCapU}, the used numerical parameters are presented in Table~\ref{param}. % Note that $V_0$ ralates to cap

\begin{table}[h]
\begin{center}
\caption{Parameters used for numerical calculations.}\label{param}
\begin{tabular}{|c|c|c|c|}
\hline
%{\it symbol}&{\it physical meaning}&{\it numerical value}\rule{0mm}{5mm}\\[0.5mm]
~Parameter~& ~substrate & ~$Ta_2O_5$ layer & ~$SiO_2$ layer \rule{0mm}{5mm}\\[0.5mm]
\hline
$T$, K &\multicolumn{3}{c}{290}~\vline\\
%$w$, m &\multicolumn{3}{c}{0.12}~\vline\\
%$R$, m &\multicolumn{3}{c}{0.31}~\vline\\
%$H$, m &\multicolumn{3}{c}{0.30}~\vline\\
$\Lambda$, m &\multicolumn{3}{c}{$1.064\times10^{-6}$}~\vline\\
\hline
$N_\text{CM}$ &~ - & $20$ &~ $19+\text{cap}$\\
$n$ &~ 1.45 &~ 2.035 &~ 1.45\\
%$\rho$, kg/m$^3$ &~ $2202$ &~ $6850$ &~ $2202$\\
$Y$, Pa &~ $72\times10^{9}$ &~ $140\times10^{9}$ &~ $72\times10^{9}$\\
$\nu$ &~ $0.17$ &~ $0.23$ &~ $0.17$\\
$\phi$ &~ $4\times10^{-10}$ &~ $2\times10^{-4}$ &~ $4\times10^{-5}$\\
\hline
\end{tabular}
\end{center}
\end{table}

Now we may compare the estimate of  accurate formula for spectral density $S_X$ of Brownian noise  with approximation $S_{\zeta_c}$ used previously \cite{02Harry,06Harry}. Numerical calculations for parameters presented in Table~\ref{param} gives:
\begin{align}
\label{estSX}
 \frac{S_{\zeta_c}- S_X}{S_X}\simeq 0.0539
\end{align}
We see that accurate calculation gives slightly less value of spectral density. However, the difference is quite modest --- about $5$ percents only.  We may qualitatively explain it by the fact that the tagent strains $u_{rr},\ u_{\phi\phi}$ are the same for all layers and they make considerable contribution into elastic energy. The  account of different coefficients $\epsilon_i$ change only strain $u_{zz}$, which is smaller than tagent strains for the majority layers.

\subsection{Mirror as a finite cylinder} 
We repeated our calculations of coating Brownian noise for mirror as a finite cylinder  using results of \cite{98Bondu,00Thorne,09Somiya}. For estimates we used the parameters listed in Table~\ref{param}. For fused silica cylinder planned in third generation laser gravitational antenna (Einstein Telescope, radius $R=0.30$~m, height $H=0.31$~m, $w=0.12$~m)  we have got 
\begin{align}
\label{estSzc}
 \left.\frac{S_{\zeta_c}- S_X}{S_X}\right|_\text{ET}\simeq 0.0524
\end{align}
For cylindric test mass planned in Advanced LIGO (radius $R=0.17$~m, height $H=0.20$~m, $w=0.06$~m) we have got
\begin{align}
\label{estSzcLIGO}
 \left.\frac{S_{\zeta_c}- S_X}{S_X}\right|_\text{aLIGO}\simeq 0.0569
\end{align}
We see that the difference is practically the same as for the model of mirror as an infinite half space.

\subsection{Double mirror (Khalili etalon)}

The number of layers used in conventional coating is large enough (about $40$) --- it is the reason why explicit and approximate formulas give close numerical results as estimates (\ref{estSX}, \ref{estSzc}, \ref{estSzcLIGO}) show. However, explicit formulas may give considerably different results in double mirror \cite{04PLAa1Kh}. Recall, in conventional mirror the fluctuations of thickness of each layer in the coating are transformed into phase fluctuations of reflected wave in two ways. First, each layer makes contribution into variation of front position of coating (position of first layer) and each layer makes approximately equal contribution. (It is worth underlying that we are interested in position of front surface of mirror relative to its center of the mass.) Second, the fluctuations of the layer thickness vary the pathlength of light traveling inside it and only several first layers make the main contribution whereas inner layers make exponentially small contribution. The first effect is much bigger than the second and usually only it is taken into account. 

The idea of double mirror, put forward by F.~Khalili \cite{04PLAa1Kh} (now the term ``Khalili etalon'' is frequently used), is to displace the part of layers from the front surface to the rear surface of the mirror. In this case the thickness fluctuations of  layers on the rear surface do not make contribution into fluctuations of front surface relative to mirror's center of the mass. And number of layers on the front surface may be smaller than total number of layers. Obviously, the explicit calculation of Brownian noise of the front layers have to give considerably less value of the spectral density as compared to approximate one due to relatively small number of layers. Obviously, there are precisely layers on front surface which make the main contribution into the coating Brownian noise of double mirror. 

As an example, we have applied approach presented in this paper for  Khalili etalon in order to estimate difference between two approaches. We assume that the  coating on front surface  consists of $3$ layers of $Ta_2O_5$ and $2$ layers of $SiO_2$ plus cap (also manufactured from fused silica) and coating on rear surface --- $17$ layers of $Ta_2O_5$ and $17$ layers of $SiO_2$. The other parameters were taken the same. One may calculate coating Brownian noise in Khalili etalon by two ways. In traditional (approximate) approach one has to apply corresponding forces $\varepsilon_fF_0$ and $\varepsilon_rF_0$ forces to front and rear surfaces of cylindric mirror (the coefficients $\varepsilon_f$ and $\varepsilon_r$ are calculated from effective tansmittances of corresponding coatings), to calculate the power dissipated in coating  and, finally, to calculate spectral density $S_\text{app}(\omega)$ using formula (\ref{FDT}). We have also calculated spectral density $S_\text{acc}(\omega)$ using our approach through coefficients $\epsilon_i$. We have found that $S_\text{acc}(\omega)$ is smaller than $S_\text{app}(\omega)$ by about $17\%$:
\begin{align}
 \frac{S_\text{app}(\omega)-S_\text{acc}(\omega)}{S_\text{acc}(\omega)}\simeq 0.172
\end{align}
Here we used the parameters from Table~\ref{param} and Advanced LIGO (radius $R=0.17$~m, height $H=0.20$~m, $w=0.06$~m).

\section{Conclusion}\label{Diss}

The formulas (\ref{epsiloncap}) may be applied to the {\em explicit} calculation of Brownian, thermoelastic, thermorefractive noise in the coating. These formulas are especially important for the thermal noise {\em compensation} first proposed by J. Kimble \cite{08Kimble} and then demonstarted for thermoelastic and thermorefractive noises  \cite{08Evans, 08Gorm}. Recall that in \cite{08Evans, 08Gorm} consideration was based on approximation (\ref{phiApp}) but we hope that  formulas (\ref{epsiloncap}) will allow to formulate {\em explicit} recommendation for thickness of each layer. 

In order to do it we have to rewrite formula for reflected wave phase $\phi$ in form useful for the calculation of thermorefractive noise. For this case we have to assume that layers positions are not fluctuated $\zeta_j=0$ but there are fluctuations of light path lengths due to variation of refractive index in each layer --- see corresponding formulas (\ref{phiDelta}) in Appendix~\ref{AppA}. This formula does not account influence of cap because in accordance to recommendation of \cite{08Gorm} the thickness of additional cap layer is a subject for optimization.

 Recently non-quarter wavelength coating was suggested by I.~Pinto \cite{Pinto}. As loss angle in $Ta_2O_5$ is much bigger than in $SiO_2$ one can use, for example, $1/8$ wavelength layer of $Ta_2O_5$ and $3/8$ wavelength layer of $SiO_2$ in order to reduce  thermal coating noise. We plan to apply our approach for detail analysis of such coatings.

 We hope that proposed approach will be useful for detailed analysis of coating noise.

\acknowledgments
S.~Vyatchanin would like to thank  Ya.~Chen for fruitful discussions during visit to Caltech. We are also grateful to  D.~Heinert, S.~Hild, R.~Nawrodt and K.~Somiya for stimulating discussions about double mirror. This work was supported by LIGO team from Caltech and in part by NSF and Caltech grant PHY-0651036 and  grant 08-02-00580 from Russian Foundation for Basic Research.

\appendix

\section{Calculations of reflected wave phase}\label{AppA}

\begin{figure}[ht]
 \psfrag{rho1}{$\rho_1$}
 \psfrag{rho2}{$\rho_2$}
 \psfrag{z}{$z$} \psfrag{z1}{$z_1$}
 \psfrag{z1+}{$z_1+\zeta_1$} \
 \psfrag{Z}{$Z_2^{(0)}$}
 \psfrag{rho0}{$\rho_o$}
 \psfrag{rn}{$\rho_o=\frac {1}{n_o}$}
 \includegraphics[width=0.35\textwidth]{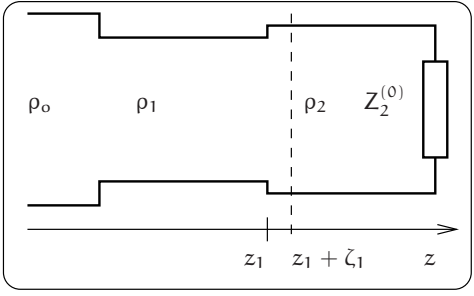}
 \caption{For calculation of dependence on displacement $\zeta_1$.}\label{zeta1fig}
\end{figure}

Here we present detailed derivation of formulas (\ref{epsiloncap}) starting with taking into account of displacement $\zeta_1$ assuming that all others are zero: $\zeta_{j\ne 1}=0$. Hence, we may consider the unperturbed piece of transmission line between $z_2\dots z_{N+1}$ as a single impedance $Z_{2}^{(0)}$ --- see Fig.~\ref{zeta1fig}. Again, the displacement $\zeta_1$ influences in two ways: a) variation of optical pathlength  between $z_0$ and $z_1$ and b) between $z_1$ and $z_2$. We consider each of them separately. 

{\em a).} Note that influence of $\zeta_1$ on pathlength between $z_0$ and $z_1$ we have almost calculated above. Indeed, calculating influence of $\zeta_0$ we have taken into account variation of exponent $\theta_1$ and expanded $\theta_1^2 \simeq -\big(1 +2ik_1(\zeta_1-\zeta_0)\big)$ 
keeping only term $\sim \zeta_0$. Now we may account displacement $\zeta_1$ just making substitution $\zeta_0\to -\zeta_1$ in (\ref{zeta0})
 \begin{align}
 \label{zeta1a}
  Z_0 &= Z_0^{(0)} - ik\zeta_1 \rho_0\left(1-\alpha_N^2\right).
 \end{align}

{\em b).} We calculate variation of $Z_1$ by $\zeta_1$ and then recalculate it into $Z_0$:
\begin{align}
R_{2}^{(0)} &= \frac{Z_{2}^{(0)}-\rho_2}{Z_{2}^{(0)}+\rho_2}=-\,\frac{1-\alpha_{N-1}}{1+\alpha_{N-1}},\quad
	\alpha_{N-1}\equiv \frac{\rho_o}{\rho_2}\left(\frac{\rho_1}{\rho_2}\right)^{N-1},
	\nonumber\\
 Z_1 &=\rho_2\frac{1 +R_{2}\theta_{2}^2}{1 -R_{2}\theta_{2}^2},\quad
 	\theta_{2}^2\simeq -(1-2ik_2\zeta_1),\nonumber\\
 Z_1 &\simeq \frac{\rho_2}{\alpha_{N-1}}
  	\left(1-\frac{ik_2\zeta_1(1-\alpha_{N-1}^2}{\alpha_{N-1}}\right), \nonumber\\
 Z_0 & =\frac{\rho_1^2}{Z_1}= Z_0^{(0)}+ ik_2\zeta_1(1-\alpha_{N-1}^2)\frac{\rho_1^2}{\rho_2}=\nonumber\\
 \label{zeta1b}
 	& =	Z_0^{(0)}+ ik\rho_o\zeta_1(1-\alpha_{N-1}^2)\, \frac{\rho_1^2}{\rho_2^2}\,.
\end{align}
Collecting (\ref{zeta1a}) and (\ref{zeta1b}) we obtain the total contribution of $\zeta_1$ in impedance $Z_0$
\begin{align}
\label{zeta1fin}
 Z_0 & =Z_0^{(0)}+ \rho_o\,ik\zeta_1\,
 	\left(\frac{\rho_1^2-\rho_2^2}{\rho_2^2}\right)
\end{align}

\begin{figure}[ht]
 \psfrag{rho1}{$\rho_1$}
 \psfrag{rho2}{$\rho_2$}
 \psfrag{z}{$z$} \psfrag{z2}{$z_2$}
 \psfrag{z2+}{$z_2+\zeta_2$} 
 \psfrag{Z}{$Z_3^{(0)}$}
 \psfrag{rho0}{$\rho_o$}
 %\psfrag{rn}{$\rho_o=\frac {1}{n_o}$}
 \includegraphics[width=0.35\textwidth]{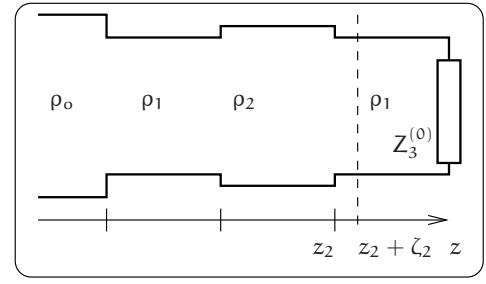}
 \caption{For calculation of dependence on displacement $\zeta_2$.}\label{zeta2fig}
\end{figure} 

Now we take into account only displacement $\zeta_2$ assuming that all others are zero: $\zeta_{j\ne 2}=0$. Hence, we may consider the unperturbed piece of transmission line between $z_3\dots z_{N+1}$ as a single impedance $Z_{3}^{(0)}$ --- see Fig.~\ref{zeta2fig}. Again we note that displacement $\zeta_2$ influences in two ways:  variation of optical pathlength a) between $z_1$ and $z_2$ and b) between $z_2$ and $z_3$. 

{\em a).} We may account displacement $\zeta_2$ just making substitution $\zeta_1\to -\zeta_2$ in (\ref{zeta1b}):
\begin{align}
\label{zeta2a}
	Z_0 &=  Z_0^{(0)}- ik\rho_o\zeta_2(1-\alpha_{N-1}^2)\, \frac{\rho_1^2}{\rho_2^2}\,.
\end{align}

{\em b).}  We calculate variation of $Z_2$ by $\zeta_2$ and then recalculate it into $Z_0$:
\begin{align}
Z_3^{(0)}  &= \frac{\rho_2^2}{Z_4^{(0)}}= \frac{\rho_2^2}{\rho_o}\left(\frac{\rho_1}{\rho_2}\right)^{N-3}\gg \rho_o,\nonumber\\
R_{3}^{(0)} &= \frac{Z_{3}^{(0)}-\rho_1}{Z_{3}^{(0)}+\rho_1}=\frac{1-\alpha_{N-2}}{1+\alpha_{N-2}},\quad
	\alpha_{N-2}\equiv \frac{\rho_o}{\rho_2}\left(\frac{\rho_1}{\rho_2}\right)^{N-2},
	\nonumber\\
 Z_2 &=\rho_1\frac{1 +R_{3}\theta_{3}^2}{1 -R_{3}\theta_{3}^2},\quad
 	\theta_{3}^2\simeq -(1-2ik_1\zeta_3),\nonumber\\
 Z_2 &\simeq \rho_1\,\frac{\alpha_{N-2}+(1-\alpha_{N-2})ik_1\zeta_2}{1+(1-\alpha_{N-2})ik_1\zeta_2}
 	\simeq\nonumber\\
 &\simeq \rho_1\alpha_{N-2} +\rho_o\,ik\zeta_2(1-\alpha_{N-2}^2), \\
 \label{zeta2b}
 Z_0 & =\frac{\rho_1^2}{\rho_2^2} Z_2= 
 	Z_0^{(0)}+ \rho_o\,ik\zeta_2(1-\alpha_{N-2}^2)\frac{\rho_1^2}{\rho_2^2}
 	\,.
\end{align}
Collecting (\ref{zeta2a}) and (\ref{zeta2b}) we obtain the total contribution of $\zeta_2$ in impedance $Z_0$
\begin{align}
 Z_0 & =Z_0^{(0)}+ 
 	\rho_o\,ik\zeta_2\,(\alpha_{N-1}^2-\alpha_{N-2}^2)\, 
 	\left(\frac{\rho_1^2}{\rho_2^2}\right)=\nonumber\\
 \label{zeta2fin}
 	&=Z_0^{(0)}+ 
 	\rho_o\,ik\zeta_2\,	\left(\frac{\rho_o^2}{\rho_2^2}\right)
 	\left(\frac{\rho_1}{\rho_2}\right)^{2N-2}\, \frac{\rho_1^2-\rho_2^2}{\rho_2^2}
\end{align}

By the same consideration we may take into account displacements $\zeta_3$ and $\zeta_4$:
\begin{align}
\label{zeta3fin}
 Z_0 & =Z_0^{(0)}+ \rho_o\,ik\zeta_3\,\left(\frac{\rho_1}{\rho_2}\right)^2 
 	\left(\frac{\rho_1^2-\rho_2^2}{\rho_2^2}\right)+\\
 	&\qquad +\rho_o\,ik\zeta_4\, \frac{(\rho_1^2-\rho_2^2)}{\rho_2^2}
 	\left(\frac{\rho_0}{\rho_2}\right)^2 
 	\left(\frac{\rho_1}{\rho_2}\right)^{2N-4}
\end{align}
and write down the final formula for effective impedance of transmission line as a function of small displacements $\zeta_i$:
\begin{subequations}
\label{Z0fin}
 \begin{align}
 Z_0 &= \rho_0 \left(\frac{\rho_1}{\rho_2}\right)^{N+1}  
 	+ik\zeta_0 \rho_0\left(1-\left[
 		\frac{\rho_o}{\rho_2}\left(\frac{\rho_1}{\rho_2}\right)^{N}\right]^2\right) + \nonumber\\ 
 	&\quad +\rho_o\,ik\zeta_1\,
 	\left(\frac{\rho_1^2-\rho_2^2}{\rho_2^2}\right)+\\
 	&\quad +\rho_o\,ik\zeta_2\,	\frac{\rho_1^2-\rho_2^2}{\rho_2^2}
 		\left(\frac{\rho_o^2}{\rho_2^2}\right)
 	\left(\frac{\rho_1}{\rho_2}\right)^{2N-2}\,+\\
 	&\quad + \rho_o\,ik\zeta_3\,\left(\frac{\rho_1}{\rho_2}\right)^2 
 	\left(\frac{\rho_1^2-\rho_2^2}{\rho_2^2}\right)+\\
 	&\quad +\rho_o\,ik\zeta_4\, \frac{(\rho_1^2-\rho_2^2)}{\rho_2^2}
 	\left(\frac{\rho_0}{\rho_2}\right)^2 
 	\left(\frac{\rho_1}{\rho_2}\right)^{2N-4}+\dots \\
 	&\quad + \rho_o\,ik\zeta_{2m-1}\,\left(\frac{\rho_1}{\rho_2}\right)^{2 m-2}
 	\left(\frac{\rho_1^2-\rho_2^2}{\rho_2^2}\right)+\\
 	&\quad +\rho_o\,ik\zeta_2m\, \frac{(\rho_1^2-\rho_2^2)}{\rho_2^2}
 	\left(\frac{\rho_0}{\rho_2}\right)^{2} 
 	\left(\frac{\rho_1}{\rho_2}\right)^{2N-2m}+\dots\nonumber	
\end{align}
\end{subequations}

\begin{figure}[ht]
 \psfrag{rho1}{$\rho_1$}
 \psfrag{rho2}{$\rho_2$}
 \psfrag{z}{$z$} \psfrag{z0}{$z_0$} \psfrag{zc}{$z_c$}
 \psfrag{zc+}{$z_c+\zeta_c$} 
 \psfrag{Z}{$Z_0$}
 \psfrag{rho0}{$\rho_o$}
 \psfrag{lambda}{$\lambda_2/2$}
 \includegraphics[width=0.25\textwidth]{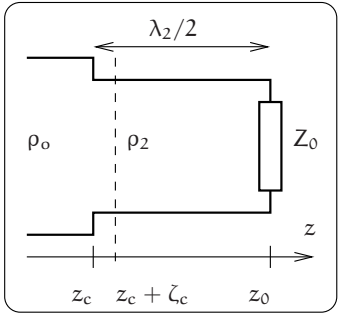}
 \caption{For calculation of equivalent impedance of the coating with a half wavelength cap.}\label{cup}
\end{figure} 

The formula for the phase $\phi$ of reflected wave complex amplitude $B$  (analog of (\ref{refl})) is more informative one:
\begin{subequations}
 \begin{align}
 B &= \frac{Z_0-\rho_o}{Z_0+\rho_o}\, A e^{2ik\zeta_0}\simeq R_0^{(0)}A(1 +i\phi),\\
 i\phi&= 2ik\sum_{j=0}^{N+1}\epsilon_j\, \zeta_j.
\end{align}
\end{subequations}
One may rewrite these formulas through refractive indices ($\rho_1=1/n_1,\ \rho_2=1/n_2$, $n_1>n_2$):
\begin{subequations}
\label{phi}
 \begin{align}
 i\phi&= 2ik\sum_{j=0}^{N+1}\epsilon_j\, \zeta_j,\\
 \epsilon_0& = 
 		n_2^2\left(\frac{n_2}{n_1}\right)^{2N}, \\ 
 \epsilon_1& = 
 	\left(\frac{n_1^2-n_2^2}{n_1^2}\right),\\
 \epsilon_2& = \left(\frac{n_1^2-n_2^2}{n_1^2} \right)n_2^2
 	\left(\frac{n_2}{n_1}\right)^{2N-2},\\
 \epsilon_3& = \left(\frac{n_1^2-n_2^2}{n_1^2}\right)
 	\left(\frac{n_2}{n_1}\right)^2 ,\\
 \epsilon_4& =  \left(\frac{n_1^2-n_2^2}{n_1^2}\right) n_2^2 
 	\left(\frac{n_2}{n_1}\right)^{2N-4},\dots \\
 \epsilon_{2m-1}& = \left(\frac{n_1^2-n_2^2}{n_1^2}\right)
 	\left(\frac{n_2}{n_1}\right)^{2 m-2},\nonumber\\
 \epsilon_{2m}& =  \left(\frac{n_1^2-n_2^2}{n_1^2}\right) n_2^{2} 
 	\left(\frac{n_2}{n_1}\right)^{2N-2m},\dots\\
 \epsilon_{N+1}& =  \left(\frac{n_1^2-n_2^2}{n_1^2}\right) n_2^{2} 
 	\left(\frac{n_2}{n_1}\right)^{N-1}.
\end{align}
\end{subequations}
We see that coefficients $\epsilon_{2m}$ with {\em even} index are relatively small and their contribution may be omitted. The main contribution is made by coefficients $\epsilon_{2m-1}$ with {\em odd} subscripts --- see plot on Fig.~\ref{Epsilon}.

\begin{figure}[ht]
 \psfrag{e}{$\epsilon_j$}
 \psfrag{rho2}{$\rho_2$}
 \psfrag{z}{$z$} \psfrag{z4}{$z_4$}
 \psfrag{z4+}{$z_4+\zeta_2$} 
 \psfrag{Z}{$Z_5^{(0)}$}
 \psfrag{rho0}{$\rho_o$}
 %\psfrag{rn}{$\rho_o=\frac {1}{n_o}$}
 \includegraphics[height=0.3\textwidth,width=0.4\textwidth]{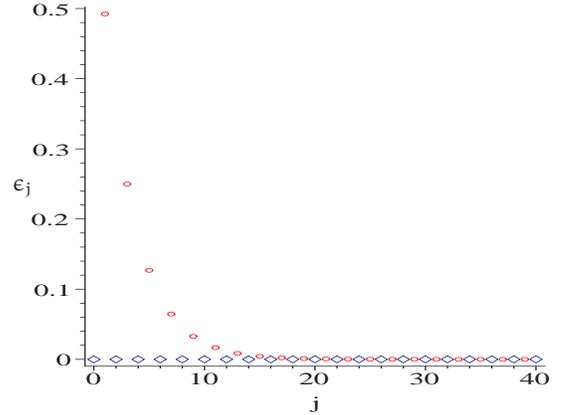}
 \caption{The coefficients $\epsilon_j$ from formulas (\ref{phi}). Red circles correspond to odd subscript and blue diamonds -- to even ones. $n_1=2.035\ (Ta_2O_5), n_2=1.45 \ (SiO_2)$. Number of $Ta_2O_5$ layers is assumed to be $N_1=20$, and number of $SiO_2$ layers --- $N_2=19$.} \label{Epsilon}
\end{figure} 

\begin{figure}[ht]
 \psfrag{ve}{$\varepsilon_j$}
 \psfrag{rho2}{$\rho_2$}
 \psfrag{z}{$z$} \psfrag{z4}{$z_4$}
 \psfrag{z4+}{$z_4+\zeta_2$} 
 \psfrag{Z}{$Z_5^{(0)}$}
 \psfrag{rho0}{$\rho_o$}
 %\psfrag{rn}{$\rho_o=\frac {1}{n_o}$}
 \includegraphics[height=0.3\textwidth,width=0.4\textwidth]{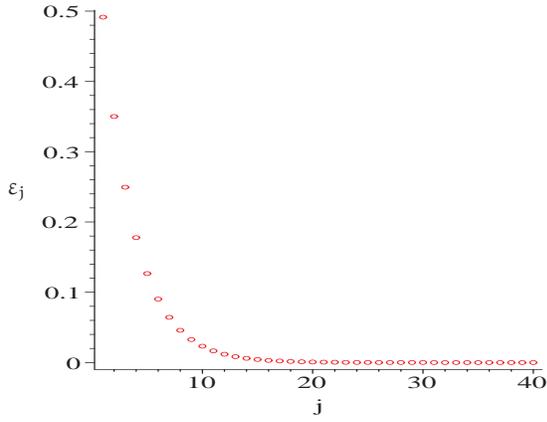}
 \caption{The coefficients $\varepsilon_j$ from formulas (\ref{phiDelta}).  $n_1=2.035\ (Ta_2O_5), n_2=1.45 \ (SiO_2)$. Number of $Ta_2O_5$ layers is assumed to be $N_1=20$, and number of $SiO_2$ layers --- $N_2=19$.}\label{varEpsilon}
\end{figure} 

We may rewrite formulas (\ref{phi}) for the reflected wave phase $\phi$ in the form useful for the calculation of thermorefractive noise. For this case we have to assume that layers positions are not fluctuated $\zeta_j=0$ but there are fluctuations of path lengths due to variation of refractive index in each layer:  
\begin{align}
\theta_j^2 & = \exp\big[2ik(n_{1,2} +\delta n_{1,2})(z_{j+1}-z_j)\big]\simeq\nonumber\\
	&\simeq 	-\left[1 + \pi \Delta_j\right],\quad 
		\Delta_{j}=\frac{\delta n_{j}}{n_{j} }.
\end{align}
Here $\Delta_j$ is the {\em relative} fluctuation of refraction index in $j$-th layer. Finally the phase of reflected wave has the following dependence on fluctuations of refractive indexes:
\begin{subequations}
\label{phiDelta}
 \begin{align}
 i\phi&= i\pi\sum_{j=1}^{N+1}\varepsilon_j\, \Delta_j,\quad \Delta_j\equiv \frac{\delta n_j}{n_j},\\
 \varepsilon_1& = \frac{1}{n_1}\left(1-	n_2^2\left(\frac{n_2}{n_1}\right)^{2N}\right), \\ 
 \varepsilon_2& =\frac{1}{n_1}\left(\frac{n_2}{n_1}\right)
 	\left(1-	n_2^2\left(\frac{n_2}{n_1}\right)^{2(N-1)}\right),\\
 \varepsilon_3& = \frac{1}{n_1}\left(\frac{n_2}{n_1}\right)^2
 	\left(1-	n_2^2\left(\frac{n_2}{n_1}\right)^{2(N-2)}\right),\\
 \varepsilon_4& = \frac{1}{n_1}\left(\frac{n_2}{n_1}\right)^3
 	\left(1-	n_2^2\left(\frac{n_2}{n_1}\right)^{2(N-3)}\right),\ \dots\ ,\\
 \varepsilon_N & = \frac{1}{n_1}\left(\frac{n_2}{n_1}\right)^{N-1}
 	\left(1-	n_2^2\left(\frac{n_2}{n_1}\right)^{2}\right), \\
 \varepsilon_{N+1}& =  \frac{1}{n_1}\left(\frac{n_2}{n_1}\right)^N
 	\left(1-	n_2^2\left(\frac{n_2}{n_1}\right)^{0}\right).
\end{align} 
\end{subequations}
We present plot $\varepsilon_j$ in Fig.~\ref{varEpsilon}.

\paragraph*{Account of a cap.}
To account for the outer half wavelength layer (cap) we have to consider the transmission line model  shown in Fig.~\ref{cup}. In the zeroth approximation we put $\zeta_c=\zeta_0=0$ and obtain
$ Z_c^{(0)}= Z_0^{(0)}$. In the first order approximation we use the expansion in series  for $\theta_0$ to account  for layer positions fluctuations:
\begin{align}
\theta_0^2 &\simeq 1+ 2ik_2(\zeta_0-\zeta_c),\\
 Z_c &\simeq \rho_2 \frac{2Z_0+(Z_0^{(0)}-\rho_2)2ik_2(\zeta_0-\zeta_c)}{
 	2\rho_2-(Z_0^{(0)}-\rho_2)2ik_2(\zeta_0-\zeta_c)}=\nonumber\\
 	&= Z_0 + ik_2(\xi_0-\xi_c)\left(\frac{(Z_0^{(0)})^2-\rho_2^2}{\rho_2}\right)
\end{align}
Now using (\ref{Z0fin}) and $k_{1,2}=k\rho_o/\rho_{1,2}$ we rewrite:
\begin{subequations}
\label{Zcfin}
 \begin{align}
 Z_c &= \rho_0 \left(\frac{\rho_1}{\rho_2}\right)^{N+1} +\\ 
 	&\quad+\rho_0 ik\zeta_c \left(1-\left[
 		\frac{\rho_o}{\rho_2}\left(\frac{\rho_1}{\rho_2}\right)^{N+1}\right]^2\right) +\\
 	&\quad +\rho_o\,ik\zeta_0\,	\frac{\rho_1^2-\rho_2^2}{\rho_2^2}
 		\left(\frac{\rho_o^2}{\rho_2^2}\right)
 	\left(\frac{\rho_1}{\rho_2}\right)^{2N}\,+\\
 	&\quad +\rho_o\,ik\zeta_1\,
 	\left(\frac{\rho_1^2-\rho_2^2}{\rho_2^2}\right)+\\
 	&\quad +\rho_o\,ik\zeta_2\,	\frac{\rho_1^2-\rho_2^2}{\rho_2^2}
 		\left(\frac{\rho_o^2}{\rho_2^2}\right)
 	\left(\frac{\rho_1}{\rho_2}\right)^{2N-2}\,+\\
 	&\quad + \rho_o\,ik\zeta_3\,\left(\frac{\rho_1}{\rho_2}\right)^2 
 	\left(\frac{\rho_1^2-\rho_2^2}{\rho_2^2}\right)+\\
 	&\quad +\rho_o\,ik\zeta_4\, \frac{(\rho_1^2-\rho_2^2)}{\rho_2^2}
 	\left(\frac{\rho_0}{\rho_2}\right)^2 
 	\left(\frac{\rho_1}{\rho_2}\right)^{2N-4}+\dots \\
 	&\quad + \rho_o\,ik\zeta_{2m-1}\,\left(\frac{\rho_1}{\rho_2}\right)^{2 m-2}
 	\left(\frac{\rho_1^2-\rho_2^2}{\rho_2^2}\right)+\\
 	&\quad +\rho_o\,ik\zeta_2m\, \frac{(\rho_1^2-\rho_2^2)}{\rho_2^2}
 	\left(\frac{\rho_0}{\rho_2}\right)^{2} 
 	\left(\frac{\rho_1}{\rho_2}\right)^{2N-2m}+\dots\nonumber	
\end{align}
\end{subequations}
From these formulas one can obtain expressions (\ref{epsiloncap}) to expand into series the reflected wave $B$
\begin{align}
 B &= \frac{Z_c-\rho_o}{Z_c+\rho_o}\, A e^{2ik\zeta_0}\simeq R_0^{(0)}A(1 +i\phi)
 \end{align}

% \begin{widetext} 
%\input{./Drafts/MultiLayerCalc.tex} 
%\input{./Drafts/LLFinite.tex} 
%\input{./Drafts/LLFiniteB.tex}
%\end{widetext}
% \input{./Drafts/DoubleM.tex}
% \appendix
%\input{./Drafts/AppInt.tex}
%\input{./Drafts/paramCM.tex}
%\input{./Drafts/bibl.tex} 
%\input{./Drafts/bibl.tex}

\end{document}